\documentclass[%
 reprint,
 amsmath,amssymb,
 aps,
 prd, 
]{revtex4-2}

\usepackage{graphicx}
\usepackage{dcolumn}
\usepackage{bm}
\usepackage{mathrsfs}
\usepackage{float}
\usepackage{hyperref}
\usepackage{xcolor}

\newcommand{\f}{\frac}
\newcommand{\n}{\nonumber}
\newcommand{\p}{\partial}

\hypersetup{
    colorlinks=true,
    linkcolor=blue,
    filecolor=magenta,
    urlcolor=blue,
    citecolor=blue,
}

\begin{document}


\title{Constraints on the cosmic reheating history from PTA data}

\author{Qi-Lin Yang} 
\affiliation{Department of Physics, College of Sciences, Northeastern University, Shenyang, Liaoning, 110819, China}

\author{Shi-Jie Wang} 
\affiliation{Department of Physics, College of Sciences, Northeastern University, Shenyang, Liaoning, 110819, China}

\author{Nan Li} \email{linan@mail.neu.edu.cn}
\affiliation{Department of Physics, College of Sciences, Northeastern University, Shenyang, Liaoning, 110819, China}

\date{\today} 

\begin{abstract}

The cosmic reheating is an important epoch in the very early Universe, and has significant impacts on the primordial black hole (PBH) abundance and the scalar-induced gravitational wave (SIGW) spectrum. The relevant energy scale and the constituents of cosmic medium are encoded in the reheating frequency $f_{\mathrm{reh}}$ and the equation of state parameter $w$. In this work, we perform a full parameter space Bayesian inference, using the NANOGrav 15-yr dataset to constrain the thermal history of reheating. Treating $f_{\mathrm{reh}}$ and $w$ as free parameters, we systematically investigate the Dirac $\delta$, narrow and broad log-normal primordial power spectra. Current pulsar timing array data can effectively constrain the peak frequency of the SIGW spectrum as $f_{\rm p}\sim 10^{-7.6}$--$10^{-7.0}$ Hz, along with $f_{\rm reh}\sim 10^{-9.0}$--$10^{-8.3}$ Hz and $w\sim 0.19$--$0.79$ (the $68\%$ confidence level), when the power spectrum has a relatively narrow width $\Delta \lesssim 1$. A broader power spectrum introduces a smoothing effect that obscures the characteristic frequency knee in the SIGW spectrum, leading to the loss of information about cosmic reheating. Furthermore, for the Dirac $\delta$ and narrow spectra, the PBH masses associated with the SIGWs are anchored at the planetary mass scale around $10^{-6}$--$10^{-4} M_{\odot}$. This establishes a consistent physical framework that naturally bridges the nanohertz gravitational wave background with the dark matter candidates implied by the Optical Gravitational Lensing Experiment microlensing events.

\end{abstract}


\maketitle


\section{Introduction}

In the very early Universe, cosmic inflation drove an almost exponential expansion of space \cite{Starobinsky:1980te, Guth:1980zm, Linde:1981mu}. While inflation elegantly solved the horizon and flatness puzzles, it simultaneously diluted all preexisting constituents. After inflation, the Universe was left in an extremely empty and cold state, with its total energy density dominated by the inflaton field. Seemingly, this contradicts the initial dense plasma state characterizing the hot big bang model, so the Universe must have undergone a transitional epoch known as cosmic reheating \cite{Kofman:1994rk, Allahverdi:2010xz, Amin:2014eta}. 

By the end of inflation, the inflaton field $\phi$ coherently oscillates around the minimum of its potential $V(\phi)$. During this epoch, the cosmic expansion rate depends on the specific shape of $V(\phi)$. For a polynomial potential $V(\phi) \propto \phi^n$, the time-averaged equation of state (EOS) parameter is approximately $w = (n-2)/(n+2)$ \cite{Turner:1983he, Ford:1986sy}. For instance, the oscillation in a harmonic potential ($n=2$) yields $w=0$, leading to an expansion history mimicking a pressureless dust-dominated Universe; conversely, a steeper potential can result in $w > 1/3$ or even approach a kinetic-energy-dominated epoch with $w \to 1$. Therefore, the parameter $w$ during the reheating phase inherently possesses significant theoretical uncertainty, spanning the range $-1/3 \le w \le 1$ \cite{Spokoiny:1993kt, Peebles:1998qn, Brax:2005uf, Hossain:2014xha}.

During the reheating era, the inflaton must transfer its energy to primary particles through different physical mechanisms \cite{Abbott:1982hn, Traschen:1990sw, Kofman:1997yn, Dufaux:2006ee}.
However, these particles initially have extremely high energies and are far from thermal equilibrium, so they must undergo microscopic interactions (e.g., high-energy scattering and annihilation) to redistribute their energies among various Standard Model particles \cite{Micha:2004bv, Lozanov:2016hid}. Once the interaction rates exceed the cosmic expansion rate, local equilibrium is established, forming a homogeneous thermal bath \cite{Garcia:2020wiy,Giudice:2000ex}, with its temperature defining a crucial cosmological parameter: the reheating temperature $T_{\rm reh}$ (or equivalently, the reheating frequency $f_{\rm reh}$). At this point, reheating is considered to complete. 

Generally speaking, in a microscopic perspective, the approach to the radiation-dominated (RD) era is a continuous dynamical process. In a perturbative reheating scenario, the inflaton condensate and the produced radiation can coexist for a period, and the EOS parameter $w$ will approach $1/3$, whereas in preheating or other nonperturbative scenarios, the detailed time dependence of $w$ can be more modeldependent \cite{Kofman:1997yn, Bassett:2005xm, Allahverdi:2010xz, Amin:2014eta}. 
For theoretical simplicity, we assume in this work an idealized transition that occurs instantaneously at $T_{\rm reh}$. In other words, the instantaneous transition should be understood as an effective benchmark with a time-independent $w$, in which the preradiation expansion history is characterized by an averaged $w$ and an effective reheating temperature $T_{\rm reh}$ (or, equivalently, $f_{\rm reh}$). Afterward, the Universe enters the RD era with $w=1/3$ immediately.


Therefore, the details of reheating, including its duration, temperature, and EOS parameter, constitute a critical ``missing history'' in the early Universe, exerting profound impacts on current cosmological observations. Unlike electromagnetic probes blocked by the opaque primordial plasma, gravitational waves (GWs) decouple almost immediately, providing a pristine window into the cosmic reheating history. 
For example, the evolution of a small-scale perturbation can be greatly altered, if its horizon reentry occurred during the reheating phase \cite{Papanikolaou:2020qtd, Lozanov:2023rcd, Dai:2014jja, Cook:2015vqa}. Moreover, the significant variation of the background pressure not only drastically modifies the critical collapse threshold and mass function for primordial black hole (PBH) formation \cite{Harada:2013epa, Escriva:2022duf}, but also imprints a characteristic ``frequency knee'' (corresponding to the end of reheating) in the infrared tail in the scalar-induced GW (SIGW) spectrum \cite{Assadullahi:2009jc, Inomata:2019zqy, Inomata:2019ivs, Domenech:2020kqm, Domenech:2019quo}. All these ingredients motivate our present study on the thermal history of cosmic reheating.



In the very early Universe, large-amplitude primordial scalar perturbations generated during inflation may undergo gravitational collapse to form PBHs, when these perturbations reenter the horizon \cite{Zeldovich:1967lct, Hawking:1971ei, Carr:1974nx, Meszaros:1974tb, Khlopov:1985fch}. Simultaneously, they can also result in SIGWs via the second-order effects \cite{Tomita:1967wkp, Matarrese:1993zf, Ananda:2006af, Baumann:2007yr, Saito:2008jc, Bugaev:2009zh, Domenech:2021ztg}. Hence, SIGWs are generally regarded as a complementary signal for probing PBH abundance and mass distribution, which is very helpful in explaining the composition of dark matter (DM) and the gravitational microlensing events \cite{Sasaki:2018dmp, Carr:2020gox, Green:2020jor, Afzal:2024hwj}. 

For the observational constraints on PBHs and SIGWs, on the one hand, for instance, the Optical Gravitational Lensing Experiment (OGLE) project has detected an excess of ultrashort-timescale microlensing events in the Galactic bulge \cite{Mroz:2017mvf, Niikura:2019kqi}. These events are difficult to be fully explained with known stellar populations or free-floating planets, hinting at the existence of a compact object population (e.g., PBHs), with masses between $10^{-6}$ and $10^{-4} M_{\odot}$ ($M_\odot=1.99\times 10^{33}$ g is the solar mass). On the other hand, the constraints on SIGWs mainly come from the stochastic GW background. In recent years, the international Pulsar Timing Array (PTA) collaborations (including CPTA \cite{Xu:2023wog}, EPTA \cite{EPTA:2023fyk, EPTA:2023sfo, EPTA:2023xxk}, and PPTA \cite{Reardon:2023gzh, Reardon:2023zen, Zic:2023gta, Xue:2021gyq}) have successively detected potential isotropic GW background in the nanohertz frequency band \cite{InternationalPulsarTimingArray:2023mzf}. Especially, the North American Nanohertz Observatory for GWs (NANOGrav) \cite{NANOGrav:2020bcs, NANOGrav:2023gor, NANOGrav:2023hde, NANOGrav:2023hvm, NANOGrav:2023hfp} has provided evidence for an excess of the red common-spectrum signals with respect to the Hellings-Downs correlations \cite{Hellings:1983fr}. 
Primarily, supermassive black hole binaries have been considered as the possible astrophysical candidate \cite{NANOGrav:2023hfp}. However, the deviation between the fitted GW spectrum and the theoretical expectation in certain frequency bands has drawn increasing attention to cosmological origin (e.g., SIGWs) \cite{Franciolini:2023pbf, Domenech:2024rks, Inomata:2023zup, Cai:2023dls, Wang:2023ost, Liu:2023ymk, Franciolini:2023wjm, Figueroa:2023zhu, Yi:2023mbm, Wang:2025kbj}. Altogether, cross-verifying the PBH mass window detected by OGLE with the GW background observed in the PTA band will become a crucial frontier in the multimessenger astronomy era, connecting DM research and early-Universe physics.

Nevertheless, the present-day SIGW spectrum depends on many aspects. In addition to the shape and amplitude of the primordial power spectrum, it is also highly sensitive to the EOS parameter $w$ of the cosmic medium upon horizon reentry. 
As we have stated, $w$ can span a much broader range $-1/3\leq w\leq 1$, if the reheating process is more physically complex. Furthermore, another influential factor is the reheating frequency $f_{\rm reh}$. In previous studies on PTA data, $f_{\rm reh}$ was always set fixed \textit{a priori} (e.g., $f_{\rm reh}=4\times 10^{-10}$ Hz in Ref. \cite{Domenech:2024rks}). While such approach reduces computational dimensionality, it is very likely to suppress the intrinsic degeneracy in the parameter space, leading to incorrect inferences of physical parameters, such as the PBH mass $M$ and the peak frequency $f_{\rm p}$ of the SIGW spectrum. 

With all these physical considerations taken into account, we will comprehensively study the influences of the reheating history on PBH and SIGW physics, allowing the reheating frequency $f_{\rm reh}$ and the EOS parameter $w$ to be free model parameters. The impacts on the SIGW parameter estimations from the uncertainties in reheating history are systematically investigated, by means of a full parameter space Bayesian inference approach, combined with the NANOGrav 15-yr dataset. For comparison, three representative types of primordial power spectra (i.e., the Dirac $\delta$, narrow log-normal, and broad log-normal spectra) are explored. We specifically choose these three models because they can cover the most general cases of spectral widths. Comparing these spectra allows us to test whether the characteristic features of the cosmic reheating history are robustly preserved or artificially smeared out by the assumed spectral shape, which is crucial for establishing reliable physical connections between PTA signals and PBH physics.

This paper is organized as follows. In Sec. \ref{sec:theory}, we briefly review the calculations of the PBH mass, abundance, and SIGW spectrum. Three types of primordial power spectra are also introduced. In Sec. \ref{sec:analysis}, various model parameters for the cosmic reheating history are constrained in detail, via Bayesian analysis and the NANOGrav 15-yr dataset. We conclude in Sec. \ref{sec:con}. We work in the natural system of units and set $c=\hbar=k_{\rm B}=1$.

\section{PBHs and SIGWs} \label{sec:theory}

In this section, we briefly review the cosmological perturbations and primordial power spectrum, and then calculate the PBH mass, abundance, and primordial and present-day SIGW spectra, respectively. 

\subsection{Cosmological perturbations and primordial power spectrum} \label{sec:inf}

We start from the single-field inflation model, with the action as
\begin{align}
S=\int{\rm d}^4x\,\sqrt{-g}\left[\f{m_{\rm P}^2}{2} R - \f{1}{2}\p_\mu\phi\p^\mu\phi-V(\phi)\right], \label{S}
\end{align}
where $R$ is the Ricci scalar, $\phi$ is the inflaton field, $V(\phi)$ is its potential, and $m_{\rm P}=1/\sqrt{8\pi G}$ is the reduced Planck mass, respectively. 
From Eq. (\ref{S}), the effective pressure and energy density of the inflaton field read $p=\phi^{\prime 2}/(2a^2) - V$ and $\rho=\phi^{\prime 2}/(2a^2) + V$, where $a$ is the scale factor and a prime denotes the derivative with respect to the conformal time $\tau$. Then, the EOS parameter $w$ is defined as 
\begin{align}
w\equiv\f{p}{\rho}=\f{1}{3}\f{1-\beta}{1+\beta}, \n
\end{align}
where a new parameter $\beta$ is introduced for convenience, showing the derivation of $w$ from $1/3$ in the RD era, 
\begin{align}
\beta \equiv \f{1-3w}{1+3w}. \n
\end{align}
Assuming the background Universe is described by the flat Friedmann-Lema\^{\i}tre-Robertson-Walker metric, the evolution of $a(\tau)$ is $a(\tau)\propto \tau^{1+\beta}$, and the conformal Hubble expansion rate is ${\cal H} \equiv a^{\prime}/a = (1+\beta)/\tau$. 

Now, we turn to the perturbations on the background metric. In Newtonian gauge, the perturbed metric (including the scalar and tensor perturbations) reads
\begin{align}
{\rm d}s^2 = a^2(\tau) \{ -(1+2\Psi) \, {\rm d}\tau^2 + [(1 + 2\Phi)\delta_{ij} + h_{ij}] \, {\rm d}x^i {\rm d}x^j \}, \n
\end{align}
where $\Psi$ and $\Phi$ are the first-order scalar perturbations, and $h_{ij}$ is the second-order transverse and traceless tensor perturbation, respectively. In the absence of anisotropic stress, the relation $\Psi=-\Phi$ holds. 

For a given wave number $k$, the equation of motion for Newtonian potential $\Phi(\tau,k)$ in the Fourier space is
\begin{align}
\Phi^{\prime\prime} + 2\varepsilon{\cal H} \Phi^{\prime} + k^{2}\Phi = 0, \label{Phi}
\end{align}
where $\varepsilon \equiv 1 - {{\cal H}^{\prime}}/ {{\cal H}^{2}} =(2+\beta)/(1+\beta)$ is the slow-roll parameter. Solving Eq. (\ref{Phi}), while requiring $\Phi$ to approach a constant on superhorizon scales ($k \ll {\cal H}$) and to match the inflationary initial conditions, the analytical solution of $\Phi(\tau,k)$ can be obtained as \cite{Domenech:2020kqm}
\begin{align}
\Phi(\tau,k) = \Phi_{\rm i}(k) \left(\f{2}{k\tau}\right)^{{3}/{2} +\beta} \Gamma \left(\f{5}{2}+\beta \right)J_{{3}/{2}+\beta}(k\tau), \n
\end{align}
where $J_{{3}/{2}+\beta}$ is the Bessel function of the first kind, and $\Phi_{\rm i}(k) = \f{2+\beta}{3+2\beta} {\cal R}_{\rm i}(k)$ is the initial value of $\Phi(\tau,k)$, with ${\cal R}_{\rm i}(k)$ being the initial value of the conserved curvature perturbation ${\cal R}(\tau,k)$ on comoving slices.

Furthermore, the physical information of the scalar perturbations is encoded in its dimensionless primordial power spectrum ${\cal P}_{\cal R}(k)$, defined as 
\begin{align}
\left.{\cal P}_{{\cal R}}(k)\equiv\f{k^3}{2\pi^2}|{\cal R}_k|^2 \right|_{k\ll {\cal H}}, \n 
\end{align}
which is essentially the Fourier transform of the two-point correlation function of ${\cal R}$, evaluated when the scale is far beyond the horizon ($k\ll{\cal H}$). In general, ${\cal P}_{\cal R}(k)$ can be obtained from $V(\phi)$ by solving the perturbed Einstein equations. Nevertheless, this procedure is usually rather complicated, so it is more convenient to directly parametrize ${\cal P}_{\cal R}(k)$ in different specific forms.

In this work, we will explore three frequently used primordial power spectra: the Dirac $\delta$, narrow log-normal, and broad log-normal spectra, respectively. Following Refs. \cite{Cai:2018dkf, Kawasaki:1997ju, Pi:2017gih, Ozsoy:2023ryl}, we assume that ${\cal P}_{\cal R}(k)$ features a peak around a wave number $k_{\rm p}$ and parametrize it in a log-normal function as
\begin{align}
{\cal P}_{{\cal R}}(k) = \f{A}{\sqrt{2\pi}\Delta} \exp \left[-\f{\ln^2(k/k_{\rm p})}{2\Delta^{2}}\right], \n
\end{align}
where $A$ is the amplitude, and $\Delta$ is the spectral width of ${\cal P}_{\cal R}(k)$. We mean narrow or broad spectra by $\Delta$ being smaller or larger than 1. Especially, in the limit $\Delta \to 0$, the log-normal spectrum naturally reduces to the Dirac $\delta$ spectrum as
\begin{align}
{\cal P}_{\cal R}(k) = A\delta(\ln(k/k_{\rm p})). \n
\end{align}

\subsection{PBH mass and abundance}

Below, we evaluate the mass and abundance of PBHs formed during an era with a general EOS parameter $w$. PBHs are formed through gravitational collapse when the local density contrast $\delta\equiv\delta\rho/\overline{\rho}$ surpasses a certain threshold $\delta_{\rm c}$ upon horizon reentry, with the most accurate analytical expression of $\delta_{\rm c}$ being \cite{Harada:2013epa}
\begin{align}
\delta_{\rm c}(w) = \frac{3(1+w)}{5+3w} \sin^2\left(\frac{\pi\sqrt{w}}{1+3w}\right). \n
\end{align} 

For PBH formation, in the Carr-Hawking collapse model \cite{Carr:1974nx}, the PBH mass $M$ is related to the horizon mass $M_{\mathrm{H}}$ as $M=\kappa M_{\mathrm{H}} =4\pi \kappa m^2_{\text{P}}/H$, where $\kappa\approx 0.2$ is the collapse efficiency, characterizing the fraction of the horizon mass that ends up in PBHs, and $H$ is the Hubble expansion rate. Then, considering the conservation of entropy in the adiabatic cosmic expansion, we obtain
\begin{align}
\frac{M}{M_\odot} = 1.58 \left(\frac{13.5}{g_{s,{\rm reh}}}\right)^{1/6} \left(\frac{k_{\rm reh}}{k_{\rm{p}}}\right)^{\frac{3(1+w)}{1+3w}} \left(\frac{130\,\rm{MeV}}{T_{\rm{reh}}}\right)^2, \label{MM}
\end{align}
where $k_{\rm reh}$ is the comoving wave number of the mode that reenters the horizon exactly at the end of reheating, and $T_{\rm reh}$ and $g_{s,{\rm reh}}$ are the cosmic temperature and the effective number of relativistic degree of freedom for entropy density at that time.

To quantify the PBH abundance, we utilize the Press-Schechter formalism \cite{Press:1973iz}, in which the mass fraction $\beta_{\rm PBH}(M)$ of the PBHs at the time of their formation is evaluated as \cite{Sasaki:2018dmp, Domenech:2020ers}
\begin{align}
\beta_{\rm PBH}(M) = \frac{\kappa}{2} {\rm erfc} \left(\frac{\delta_{\rm c}(w)}{\sqrt{2}\sigma(M)}\right), \n
\end{align}
where $\sigma(M)$ denotes the variance of the density perturbations smoothed over the mass scale $M$,
\begin{align}
\sigma^2 (M)= 4 \left(\frac{1+w}{5+3w}\right)^2 \int W^2(kR)(kR)^4 {\cal P_R}(k) \,{\rm d}\ln k,\n
\end{align}
with $W(kR)$ being the window function (typically chosen as Gaussian form) used to smooth the perturbations over the scale $R$ associated with the PBH mass $M$. 

Next, the PBH abundance is defined as its proportion in DM today, 
\begin{align}
f_{\rm PBH}\equiv\f{\rho_{\rm PBH,0}}{\rho_{\rm DM,0}}, \n 
\end{align}
where $\rho_{\rm PBH,0}$ and $\rho_{\rm DM,0}$ are the present-day energy densities for PBH and DM. For the monochromatic PBH mass function, $f_{\mathrm{PBH}}$ is naturally proportional to $\beta_{\mathrm{PBH}}$ and can be obtained as \cite{Domenech:2020ers}
\begin{align}
f_{{\rm PBH}} & = 1.96 \times 10^{12}\left(\frac{g_{s,\rm reh}}{14.25}\right)^{-1} \left(\frac{g_{\rho,\rm reh}}{13.5}\right) \n \\
&\quad \times \left(\frac{k_{\rm p}}{k_{\rm reh}}\right)^{\frac{6w}{1+3w}} \left(\frac{T_{\rm reh}}{130\,{\rm MeV}}\right) \beta_{\rm PBH}, \label{fM}
\end{align}
where $g_{\rho,\rm reh}$ is the effective number of relativistic degree of freedom for energy density at $T_{\rm reh}$.

\subsection{Primordial SIGW spectrum} \label{sec:GW}

Below, we concisely review the derivation of the primordial SIGW spectrum with a general EOS parameter. 

First, the equation of motion for the Fourier mode $h_{\lambda}(\tau,{\bf k})$ of the tensor perturbation $h_{ij}$ reads
\cite{Kohri:2018awv, Espinosa:2018eve, Yuan:2019udt, Pi:2020otn} 
\begin{align}
h_{\lambda}^{\prime\prime} + 2{\cal H} h_{\lambda}^{\prime} + k^{2} h_{\lambda} = s_{\lambda}(\mathbf{k}), \n
\end{align}
where $\lambda$ denotes the polarization state of $h_{ij}$, and $s_{\lambda}(\mathbf{k})$ is the source term, composed of the scalar perturbations and their derivatives convoluted in momentum space,
\begin{align}
s_{\lambda}(\mathbf{k})& =8 \int \f{{\rm d}^3{q}}{(2\pi)^3} \, e_{\lambda}^{ij}(\mathbf{k}) q_{i} q_{j} \bigg\{ \Phi(q) \Phi(|\mathbf{k}-\mathbf{q}|)+ \f{1+\beta}{2+\beta} \n\\
&\quad \times\left[ \Phi(q) + \f{\Phi^{\prime}(q)}{{\cal H}} \right] \left[ \Phi(|\mathbf{k}-\mathbf{q}|) + \f{\Phi^{\prime}(|\mathbf{k}-\mathbf{q}|)}{{\cal H}} \right] \bigg\}, \n
\end{align}
with $e_{\lambda}^{ij}(\mathbf{k})$ being the polarization tensor. Using Green's function method, $h_\lambda$ can be solved analytically \cite{Kohri:2018awv}.

Then, the SIGW spectrum $\Omega_{{\rm GW}}(\tau,k)$ is defined as the GW energy density per logarithmic wave number as
\begin{align}
\Omega_{\rm GW}(\tau,k)\equiv\f{1}{\rho_{\rm c}}\f{{\rm d}\rho_{\rm GW}(\tau,k)}{{\rm d}\ln k}, \n 
\end{align}
where $\rho_{\rm c}$ and $\rho_{{\rm GW}}$ are the critical and GW energy densities in the early Universe. Moreover, in the transverse-traceless gauge, $\Omega_{{\rm GW}}$ can be reexpressed as \cite{Isaacson:1968hbi, Isaacson:1968zza}
\begin{align}
\Omega_{{\rm GW}}(\tau,k)=\f{1}{48}\left(\f{k}{{\cal H}} \right)^2\overline{{\cal P}_h(\tau,k)}. \n 
\end{align}
Above, ${\cal P}_h(\tau,k)$ is the dimensionless power spectrum of the tensor perturbation and can be further linked to the product of the primordial scalar power spectra ${\cal P}_{\cal R}(k)$ as \cite{Espinosa:2018eve, Kohri:2018awv, Domenech:2019quo}
\begin{align}
{\cal P}_{h}(\tau,k) &=8 \int_{0}^{\infty} {\rm d}v \int_{|1-v|}^{1+v} {\rm d}u\, \left[ \f{4v^{2}-(1-u^{2}+v^{2})^{2}}{4uv} \right]^{2} \n\\
& \quad \times {\cal P}_{{\cal R}}(ku) {\cal P}_{{\cal R}}(kv) \overline{I^{2}(u,v,x,\beta)}, \n
\end{align}
where three dimensionless variables are introduced for convenience: $u \equiv |\mathbf{k}-\mathbf{q}|/k$, $v \equiv q/k$, and $x \equiv k\tau$, respectively. Moreover, $\overline{I^{2}}$ is the oscillation average of the squared integral kernel function $I(u,v,x,\beta)$ \cite{Domenech:2020kqm},
\begin{align}
I(u,v,x,\beta) &=\f{\pi(2+\beta)}{3+2\beta} \left(\frac{4}{uvx}\right)^{{1}/{2}+\beta} \Gamma^{2} \left(\f{3}{2}+\beta \right) \n\\
& \quad \times [ Y_{{1}/{2}+\beta}(x) {\cal I}_{J} - J_{{1}/{2}+\beta}(x) {\cal I}_{Y} ], \label{I}
\end{align}
where $Y_{{1}/{2}+\beta}$ is the Bessel function of the second kind, and the integrals ${\cal I}_{J}$ and ${\cal I}_{Y}$ read
\begin{align}
{\cal I}_{J,Y}(u,v,x,\beta)&= \int_{0}^{x} {\rm d}\widetilde{x} \, \widetilde{x}^{{1}/{2}-\beta} \left\{ \begin{matrix} J_{{1}/{2}+\beta}(\widetilde{x}) \\ Y_{1/2+\beta}(\widetilde{x}) \end{matrix} \right\} \n\\
&\quad\times \bigg[ J_{{1}/{2}+\beta}(u\widetilde{x}) J_{{1}/{2}+\beta}(v\widetilde{x}) \n\\
&\quad + \f{2+\beta}{1+\beta} J_{{5}/{2}+\beta}(u\widetilde{x}) J_{{5}/{2}+\beta}(v\widetilde{x}) \bigg]. \label{jifen}
\end{align}

Next, we discuss the limiting cases of the kernel function $I(u, v, x, \beta)$. In the subhorizon approximation ($x=k\tau \gg 1$), the scales reenter the horizon well before reheating. In this case, we obtain 
\begin{align}
I(u,v,x\gg 1,\beta)\approx \f{C_{1,\beta}J_{1/2+\beta}(x) +C_{2,\beta}Y_{1/2+\beta}(x)}{x^{1/2+\beta}}. \label{subI}
\end{align}
After averaging over the high-frequency oscillations, the squared subhorizon kernel function can be expressed analytically as
\begin{align}
&\quad \overline{I^{2}(u,v,x\gg 1,\beta)} \n\\
&=\f{2^{1+4\beta}}{x^{2(1+\beta)}} \left( \f{2+\beta}{3+2\beta} \right)^{2} \Gamma^{4} \left(\f{3}{2} +\beta\right) \f{(1-y^{2})^{\beta}}{(uv)^{2}} \n\\
& \quad \times \left\{ \left[ P_{\beta}^{-\beta}(y) + \f{2+\beta}{1+\beta} P_{2+\beta}^{-\beta}(y) \right]^{2} \right. \n\\
& \left. \quad + \f{4}{\pi^{2}} \left[ Q_{\beta}^{-\beta}(y) + \f{2+\beta}{1+\beta} Q_{2+\beta}^{-\beta}(y) \right]^{2} \right\}, \label{sub}
\end{align}
where $P_{\beta}^{-\beta}(y)$ and $Q_{\beta}^{-\beta}(y)$ are the associated Legendre functions of the first and second kind, with $y \equiv 1-[1-(u-v)^{2}]/(2uv)$ being a dimensionless variable. On the contrary, in the superhorizon approximation ($x=k\tau \ll 1$), for the scales still outside the horizon prior to reheating, expanding around the peaked spectrum yields
\begin{align}
I(v,x\ll 1,vx\gg 1,\beta) \approx \widetilde{C}_{1,\beta} + \widetilde{C}_{2,\beta} x^{-2\beta}. \label{super}
\end{align}
The specific forms of the expansion coefficients $C_{1,\beta}$, $C_{2,\beta}$ in Eq. (\ref{subI}) and $\widetilde{C}_{1,\beta}$, $\widetilde{C}_{2,\beta}$ in Eq. (\ref{super}) are irrelevant for our following discussions and can be found in Ref. \cite{Domenech:2020kqm}. Altogether, the results in Eqs. (\ref{subI})--(\ref{super}) will be used to derive the present-day SIGW spectrum below.

\subsection{Present-day SIGW spectrum} 

To obtain the present-day SIGW spectrum $\Omega_{\rm GW,0}(k)$, we must carefully distinguish the scales that reenter the horizon before and after the reheating phase. The comoving wave number $k_{\rm reh}$ corresponds to the scale that exactly reenters the horizon at the end of reheating. Using the horizon-crossing condition $k = {\cal H} = (1+\beta)/\tau$, this scale is determined by $k_{\rm reh} \tau_{\rm reh} = 1+\beta$.

In the following, $\tau_{\rm reh}$ is regarded as the onset of the RD era. The instantaneous transition approximation is expected to be most reliable for the modes sufficiently far away from $k_{\rm reh}$,
and the modes most sensitive to the detailed finite-duration transition are those around $k_{\rm reh}$ \cite{Inomata:2019ivs, Domenech:2020kqm}. Therefore, $f_{\rm reh}$ should be interpreted as an effective reheating scale, determining the location of the characteristic frequency knee in the SIGW spectrum, rather than as a definite transition frequency. Moreover, we mention that, for the $w=0$ benchmark, the comparisons between the instantaneous \cite{Inomata:2019ivs} and gradual \cite{Inomata:2019zqy} transitions indicated that there can be a slight magnitude difference in the SIGW amplitude around $k_{\rm reh}$, suggesting that a finite-duration transition can introduce systematic uncertainty into the inferred reheating scale. In this sense, $f_{\rm reh}$ in our analysis should be understood as an effective scale with associated systematic uncertainty, instead of a definite microscopic transition frequency.

As we have assumed the instantaneous transition from the reheating to RD era at $\tau_{\rm reh}$, the tensor perturbation $h_\lambda$ and its time derivative $h^\prime_\lambda$ should be continuous at $\tau_{\rm reh}$. During the RD era, the modes propagate as free waves governed by a shifted time as $\widetilde{\tau} = \tau - \f{\beta}{1+\beta}\tau_{\rm reh}$, ensuring the continuity of the background expansion. Thus, the present-day SIGW spectrum $\Omega_{\rm GW,0}$ depends on the time when a scale reenters the horizon.

First, for the scales reentering the horizon before reheating ($k \gg k_{\rm reh}$), these modes have long begun oscillating and continue propagating as free waves in the RD era. Matching $h_\lambda$ and $h^\prime_\lambda$ at $\tau_{\rm reh}$ maps the subhorizon solutions directly into the RD free-wave solutions, which leads to \cite{Domenech:2020kqm}
\begin{align}
\Omega_{\rm GW,0}(k \gg k_{\rm reh}) = \f{1}{48} \left(\f{k}{{\cal H}}\right)^2 \overline{{\cal P}_{h}(\tau_{\rm reh}, k \gg k_{\rm reh})}. \n
\end{align}
Second, for the scales with horizon reentry after reheating ($k \ll k_{\rm reh}$), applying the same continuity conditions to match the superhorizon asymptotic approximations with the RD solutions yields the matching coefficients $C_{1,\rm RD}$ and $C_{2,\rm RD}$ in the RD era as \cite{Domenech:2020kqm}
\begin{align}
C_{1,\rm RD} &= \widetilde{C}_{1,\beta} + \f{1-\beta}{1+\beta} (k\tau_{\rm reh})^{-2\beta}\widetilde{C}_{2,\beta}, \n\\ 
C_{2,\rm RD} &= \f{2\beta}{(1+\beta)^{2}} (k\tau_{\rm reh})^{1-2\beta} \widetilde{C}_{2,\beta}. \n
\end{align}

Following the above matching procedures, the GW energy density $\rho_{\rm GW}$ on the subhorizon scales decays identically to radiation as $\rho_{\rm GW} \propto a^{-4}$, whereas the background energy density decays as $\overline{\rho} \propto a^{-3(1+w)}$. Then, using entropy conservation in the adiabatic cosmic expansion, we are able to convert the SIGW spectrum at the end of reheating $\Omega_{\rm GW,reh}(k)$ to that at present $\Omega_{\rm GW,0}(k)$ as \cite{Domenech:2024rks}
\begin{align}
\Omega_{\rm GW,0}(k)h^{2} &= 3.40 \times 10^{-5} \left( \f{g_{\rho,\rm reh}}{10.75} \right) \left( \f{g_{s,\rm reh}}{10.75} \right)^{-{4}/{3}} \n\\
&\quad \times \left( \f{\Omega_{\rm r,0} h^{2}}{2.47 \times 10^{-5}} \right) \Omega_{\rm GW,reh}(k), \n
\end{align}
where $h$ is the reduced Hubble constant, and $\Omega_{\rm r,0}$ is the present-day radiation energy density parameter \cite{Planck:2018vyg}. 


Furthermore, the reheating frequency $f_{\rm reh}$ can be related to the temperature $T_{\rm reh}$.
Using the horizon-crossing condition $k_{\rm reh}=a_{\rm reh}H_{\rm reh}$ with $f_{\rm reh}={k_{\rm reh}}/(2\pi a_0)$, the Friedmann equation at the onset of the RD era $H_{\rm reh}^2=\pi^2g_{\rho,{\rm reh}}T_{\rm reh}^4/(90m_{\rm P}^2)$, and the entropy conservation condition after reheating $g_{s,{\rm reh}}T_{\rm reh}^3a_{\rm reh}^3 =g_{s,0}T_0^3a_0^3$, we have
\begin{align}
f_{\rm reh}=\frac{1}{2\pi}\left(\frac{g_{s,0}}{g_{s,{\rm reh}}}\right)^{1/3} \left(\frac{\pi^2g_{\rho,{\rm reh}}}{90}\right)^{1/2} \frac{T_0}{m_{\rm P}}T_{\rm reh}. \n
\end{align}
Substituting detailed numerical values, we can obtain
\begin{align}
f_{\rm reh}&=1.81\times10^{-8}\left(\frac{g_{\rho,{\rm reh}}}{10.75}\right)^{1/2}\left(\frac{g_{s,{\rm reh}}}{10.75}\right)^{-1/3} \left(\frac{T_{\rm reh}}{{\rm GeV}}\right){\rm Hz}, \label{yql}
\end{align}
or equivalently,
\begin{align}
T_{\rm reh}&= 55.3\left(\frac{g_{\rho,{\rm reh}}}{10.75}\right)^{-1/2}\left(\frac{g_{s,{\rm reh}}}{10.75}\right)^{1/3} \left(\frac{f_{\rm reh}}{{\rm nHz}}\right){\rm MeV}. \label{eq:Treh-freh}
\end{align}
In this way, $T_{\rm reh}$ is directly expressed in terms of $f_{\rm reh}$.

Last, by using the conversion relation between the GW frequency $f$ and the comoving wave number $k$ today,
\begin{align}
f=1.5\times 10^{-6}\left(\frac{k}{{\rm Mpc}^{-1}}\right)~{\rm nHz}, \n
\end{align}
we can finally obtain the present-day SIGW spectrum $\Omega_{\rm GW,0}(f)h^2$ in terms of the frequency $f$.

\section{Constraints on the cosmic reheating history} \label{sec:analysis}

Having established the theoretical framework for SIGW and its dependence on reheating physics (e.g., the reheating frequency $f_{\rm reh}$ and the EOS parameter $w$), we now confront them with the current observational constraints. By performing Bayesian parameter inference, our work reflects the characteristics of primordial perturbations and cosmic reheating history. For these purposes, we utilize the publicly available \texttt{SIGWfast} \cite{Witkowski:2022mtg} and \texttt{PTArcade} \cite{Mitridate:2023oar} packages, with the ceffyl model \cite{Lamb:2023jls}.

\subsection{Priors} \label{sec:priors}

Table \ref{tab:parameters_prior} summarizes our model parameters in the full parameter space, their physical meanings, and the relevant prior settings. The choices of these priors are observationally motivated. (1) The peak frequency $f_{\rm p}$ and the reheating frequency $f_{\rm reh}$ are set within $[10^{-9}, 10^{-5}]$ Hz to adequately cover the nanohertz frequency band sensitive to PTA data. Albeit $f_{\rm reh}$ is treated as a free parameter in our analysis, in order to fit the signal within the PTA band, the condition $f_{\rm reh} \lesssim f_{\rm p}$ is necessary, ensuring that the reheating-induced spectral features remain in the observable window. (2) The EOS parameter $w$ spans from $0$ to $1$, comprehensively encompassing most typical reheating scenarios from a matter-dominated ($w=0$) to a kinetic-energy-dominated ($w=1$) Universe. (3) The amplitude prior $\log_{10} A \in [-4, 4]$ is wide enough to explore the entire parameter space from generating negligible SIGW signals to causing PBH overproduction. (4) For the spectral width, $\log_{10} \Delta \in [-3, 0]$ restricts the analysis to a phenomenological narrow peak, whereas $\log_{10} \Delta \in [0, 1]$ is specifically adopted to explore the smoothing effects of broad spectra.
\begin{table}[htbp]
\centering\caption{The model parameters in Bayesian parameter inference and their prior ranges: $f_{\rm p}$ is the peak frequency of the SIGW spectrum, $f_{\rm reh}$ is the reheating frequency, $w$ is the EOS parameter, and $A$ and $\Delta$ are the amplitude and width of the primordial power spectrum, respectively. The units of $f_{\rm p}$ and $f_{\rm reh}$ are hertz, and $w$, $A$, and $\Delta$ are dimensionless.}
\label{tab:parameters_prior}
\renewcommand{\arraystretch}{1.3} 
\begin{tabular}{c|c} 
\hline\hline
Parameter & Prior (uniform) \\
\hline
$\log_{10} f_{\rm p}$               & $[-9, -5]$ \\
$\log_{10} f_{\rm reh}$             & $[-9, -5]$ \\
$w$                                 & $[0, 1]$   \\
$\log_{10} A$                       & $[-4, 4]$  \\
$\log_{10}\Delta$ (narrow spectrum) & $[-3, 0]$  \\
$\log_{10}\Delta$ (broad spectrum)  & $[0, 1]$   \\
\hline\hline
\end{tabular}
\end{table}

Here, we note that the above priors are consistent with big bang nuclEOSynthesis (BBN). A conservative prerequisite is that the Universe is radiation dominated before BBN, namely, $T_{\rm reh}>T_{\rm BBN}\approx 4.1~{\rm MeV}$ \cite{Kawasaki:1999na, Hannestad:2004px, Ichikawa:2005vw, deSalas:2015glj, Hasegawa:2019jsa}. Considering the lower edge of our prior $f_{\rm reh}=10^{-9}~{\rm Hz}$, from Eqs. \eqref{yql} and \eqref{eq:Treh-freh}, we obtain $T_{\rm reh}> 55.3~{\rm MeV}$ for $g_{\rho,{\rm reh}}=g_{s,{\rm reh}}=10.75$. Therefore, the BBN consistency condition is naturally satisfied in the entire parameter space in our work, ensuring that the reheating scales probed by nanohertz PTA data are completely compatible with the early cosmic thermal history.

Moreover, before presenting the numerical posteriors, we briefly summarize the main analytical dependence of the SIGW spectrum $\Omega_{\rm GW,0}$ on the reheating parameters. (1) For the modes reentering the horizon before the end of reheating, $\Omega_{\rm GW,0}$ contains a factor as $(k/k_{\rm reh})^{-2\beta}$ \cite{Ananda:2006af, Baumann:2007zm, Kohri:2018awv, Domenech:2020kqm}.
This relation means that $w$ determines the spectral slope and amplitude, and $f_{\rm reh}$ fixes the position of the reheating-induced knee. (2) Within the infrared region with $k_{\rm reh} \ll k \ll k_{\rm p}$, the slope of the SIGW spectrum exhibits power-law behaviors that crucially depend on the form of the primordial power spectra. For the Dirac $\delta$ spectrum, the infrared scaling follows $\Omega_{\rm GW,0} \propto k^{n_{\rm IR}}$, with the spectral index being $n_{\rm IR} = 2 - 2|\beta|$; in contrast, for the log-normal spectra with finite widths, the momentum integration over the scalar modes will introduce an additional power of $k$, yielding $n_{\rm IR} = 3 - 2|\beta|$ \cite{Cai:2019cdl, Domenech:2020kqm}. (3) For the modes with $k < k_{\rm reh}$ that reenter the horizon after reheating, the standard transfer function in the RD era is recovered. Consequently, the transition at the end of reheating imprints a characteristic spectral knee at $f_{\rm knee} \sim f_{\rm reh}$. (4) Since the current PTA data cover only a finite frequency interval, the shifting of the spectral knee can be partially compensated by adjusting $w$ and $A$, inevitably leading to the posterior parameter degeneracy between $w$ and $f_{\rm reh}$, to be discussed below.

In Secs. \ref{sec:delta}--\ref{sec:broad}, we will consider three different types of primordial power spectrum: the Dirac $\delta$, narrow log-normal, and broad log-normal spectra, respectively. We perform the independent Markov chain Monte Carlo sampling and posterior analysis on the NANOGrav 15-yr dataset. The corresponding posterior distributions for all the model parameters are summarized in Table \ref{tab:parameters_posterior}.
\begin{table*}[htbp]
\centering
\caption{The posterior distributions ($68\%$ CL) for the model parameters, with three different types of primordial power spectrum: the Dirac $\delta$, narrow log-normal, and broad log-normal spectra, respectively.}
\label{tab:parameters_posterior}
\renewcommand{\arraystretch}{1.3} 
\begin{tabular}{c|c|c|c} 
\hline\hline
Parameter & Dirac $\delta$ & Narrow log-normal & Broad log-normal \\
\hline
$\log_{10} f_{\rm p}$   & $-5.87^{+0.85}_{-0.31}$ & $-7.28^{+0.30}_{-0.29}$ & $-6.18^{+1.10}_{-0.46}$ \\
$\log_{10} f_{\rm reh}$ & $-7.26^{+0.76}_{-0.75}$ & $-8.46^{+0.18}_{-0.51}$ & $-7.82^{+0.75}_{-1.17}$ \\
$w$                     & $0.50^{+0.25}_{-0.36}$  & $0.53^{+0.26}_{-0.34}$  & $0.58^{+0.40}_{-0.19}$  \\
$\log_{10} A$           & $-0.22^{+0.83}_{-0.93}$ & $-1.47^{+0.28}_{-0.49}$ & $-0.64^{+0.52}_{-0.51}$ \\
$\log_{10} \Delta$      & ---                     & $-0.81^{+0.81}_{-0.24}$ & $0.30^{+0.17}_{-0.12}$  \\
\hline\hline
\end{tabular}
\end{table*}

\subsection{Dirac $\delta$ spectrum} \label{sec:delta} 

\begin{figure}[htbp]
\begin{center}
\includegraphics[width=1\linewidth]{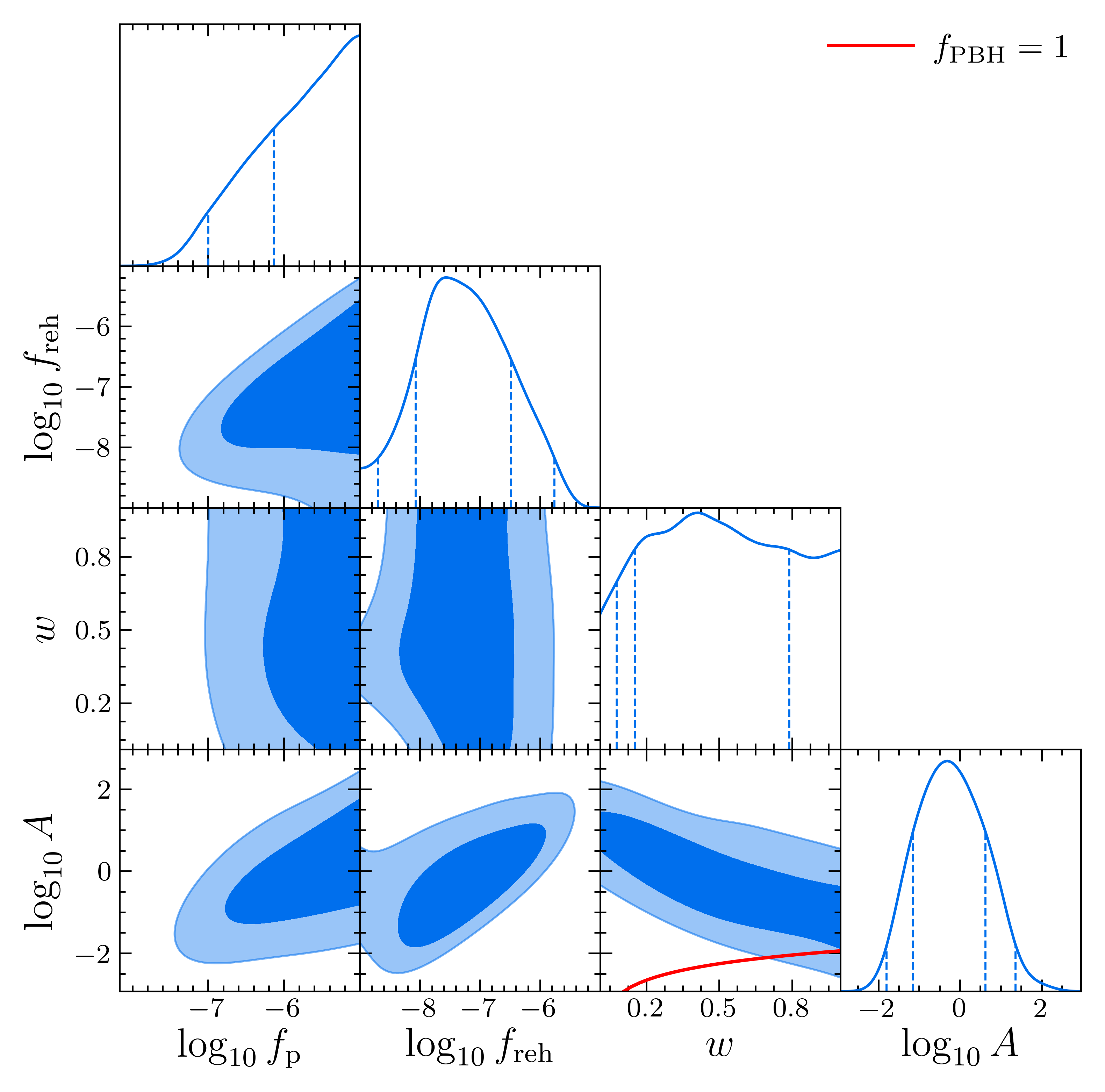}
\caption{The posterior distributions of the four model parameters $f_{\rm p}$, $f_{\rm reh}$, $w$, and $A$ for the Dirac $\delta$ spectrum, and the blue dashed lines indicate their $68\%$ and $95\%$ CLs. In the joint posterior distribution of $f_{\rm p}$ and $f_{\rm reh}$, the necessary condition $f_{\rm reh} \lesssim f_{\rm p}$ is guaranteed. In the joint posterior distribution of $w$ and $A$, the red line corresponds to the PBH abundance $f_{\rm PBH} = 1$, and only the region below this line is physically allowed to avoid PBH overproduction.} \label{fig:delta}
\end{center}
\end{figure}

We start from the simplest Dirac $\delta$ spectrum, as in this case the PBH abundance and the SIGW spectrum can be calculated analytically. The posterior distributions of the four model parameters (i.e., the peak frequency of the SIGW spectrum $f_{\rm p}$, reheating frequency $f_{\rm reh}$, EOS parameter $w$, and amplitude $A$ of the power spectrum) are shown in Fig. \ref{fig:delta}. 

First, the Dirac $\delta$ spectrum favors a higher $f_{\rm p}$, and this grants $f_{\rm reh}$ a large allowed region at high frequencies, implying that the model permits the reheating process to take place at high energies. Moreover, to preserve the power-law features in the infrared tail in the SIGW spectrum, the model requires the necessary condition $f_{\rm reh} \lesssim f_{\rm p}$, which can be clearly seen in their joint posterior distribution.

Second, for the EOS parameter $w$, it can be theoretically inferred from the slope of the infrared tail in the SIGW spectrum ($2 - 2|\beta|$ for the Dirac $\delta$ spectrum) \cite{Cai:2019cdl, Domenech:2020kqm}, but this applies for the frequencies larger than $f_{\rm reh}$. However, in our present work, $f_{\rm reh}$ is also considered as a free model parameter, and this will introduce uncertainty in the position of the characteristic cutoff in the SIGW spectrum, rendering the local slope obtained within a limited observational band unable to fully determine $w$, as can be seen in Fig. \ref{fig:delta}. 
This behavior is consistent with the relation between $\Omega_{\rm GW,0}$ and $(k/k_{\mathrm{reh}})^{-2\beta}$: changing $w$ can modify the spectral slope, whereas changing $f_{\rm reh}$ can also shift the position of the reheating-induced cutoff.

Furthermore, we discuss an important point of PBH overproduction. In the joint posterior distribution of $w$ and $A$ in Fig. \ref{fig:delta}, from Eq. (\ref{fM}), the red line illustrates the critical bound with the PBH abundance $f_{\rm PBH} = 1$, and only the region below this red line is physically allowed with $f_{\rm PBH}<1$. The highly restrictive overlap with the blue contours is located at the lower right corner, nearly outside the $95\%$ confidence level (CL) region favored by PTA data. This observation indicates the severe intrinsic tension in the Dirac $\delta$ spectrum model: the parameters that safely avoid PBH overproduction generally fail to produce intensive SIGWs in the PTA band. In other words, it is difficult to gain a balance between providing entire DM and satisfying the PTA constraints at the same time. Our result is consistent with that in Ref. \cite{Domenech:2024rks}, where it was found that, when $f_{\rm reh}$ is fixed to $4 \times 10^{-10}$ Hz, $w > 0.42$ can avoid PBH overproduction.

\begin{figure}[htbp]
\begin{center} 
\includegraphics[width=1\linewidth]{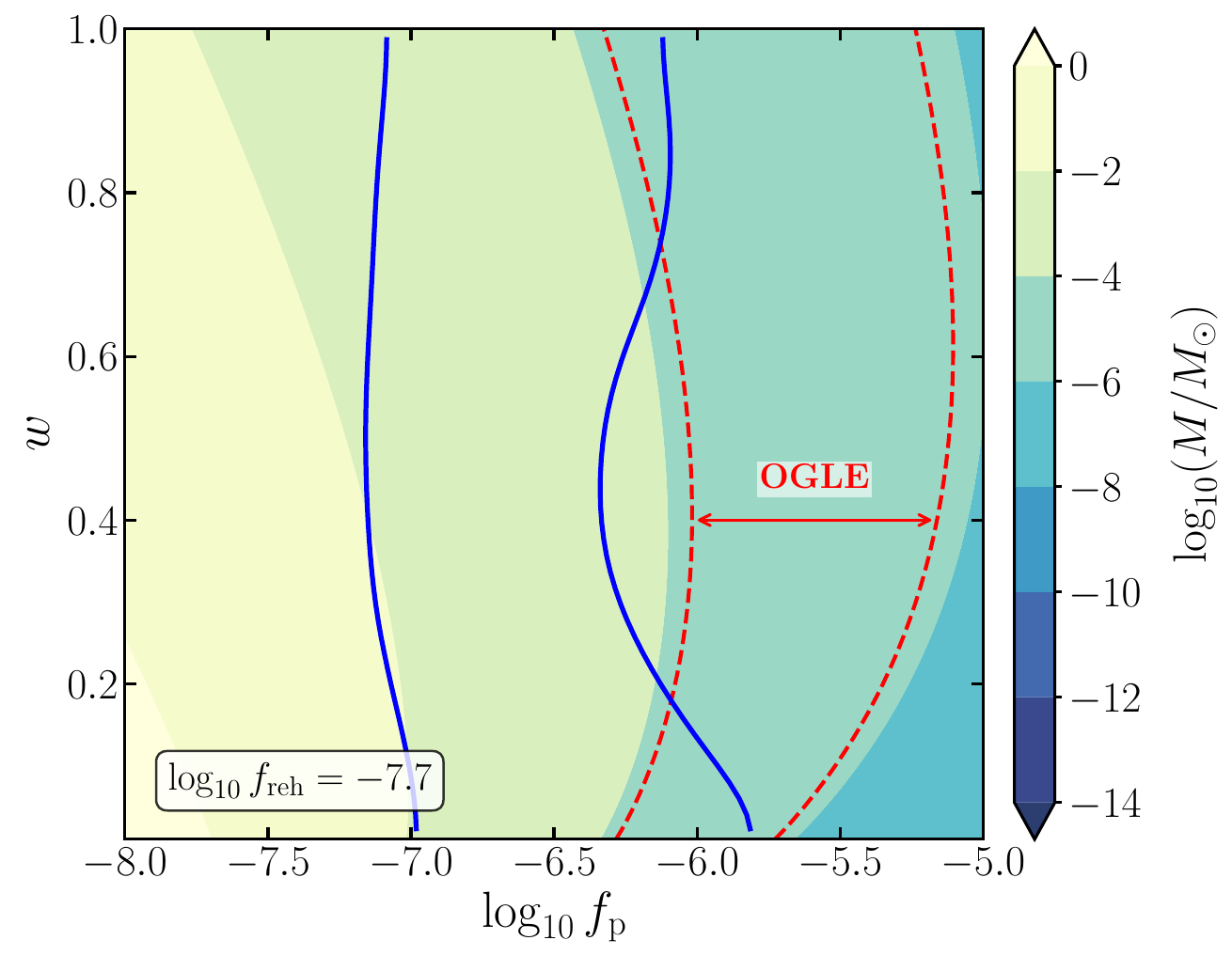}
\caption{The joint posterior distribution of the EOS parameter $w$ and the peak frequency $f_{\rm p}$ of the SIGW spectrum for the Dirac $\delta$ spectrum (same as that in Fig. \ref{fig:delta}), with the reheating frequency chosen as its maximum marginal posterior value $\log_{10}f_{\rm reh}=-7.7$. The background colors indicate different PBH masses. The blue lines indicate the allowed parameter space favored by the PTA data ($68\%$ and $95\%$ CLs), and the region between the red dashed lines corresponds to the mass range constrained by the OGLE microlensing events. Their overlap shows that, within a specific parameter window, the Dirac $\delta$ spectrum can simultaneously account for the stochastic GW background observed by the PTAs and the PBH masses constrained by the OGLE events.} \label{fig:delta_pbh}
\end{center}
\end{figure}

Last, we investigate the issue of parameter degeneracy. As $f_{\rm reh}$ is now a free model parameter, it will induce uncertainty in the characteristic cutoff position in the SIGW spectrum, leading to unavoidable parameter degeneracy between the EOS parameter $w$ and the peak frequency $f_{\rm p}$ in determining the PBH mass in Eq. (\ref{MM}). As illustrated in Fig. \ref{fig:delta_pbh}, the PTA-favored parameter space manifests as the allowed bands. Notably, a significant portion of the $68\%$ CL region intersects with the PBH mass range constrained by the OGLE microlensing events (the region between the red dashed lines). This overlap implies that, despite the parameter degeneracy introduced by releasing the thermal history of cosmic reheating, there still exists a specific parameter window, where the Dirac $\delta$ spectrum can simultaneously account for the stochastic GW background observed by the PTAs and the PBH masses corresponding to the microlensing events. 

After all, the Dirac $\delta$ spectrum is merely a simple and idealized model, and the relevant inference is of less physical essence, so we will move on to the more realistic models with finite widths in the log-normal power spectra.

\subsection{Narrow log-normal spectrum} \label{sec:narrow} 

\begin{figure*}[htbp]
\begin{center}
\includegraphics[width=0.65\linewidth]{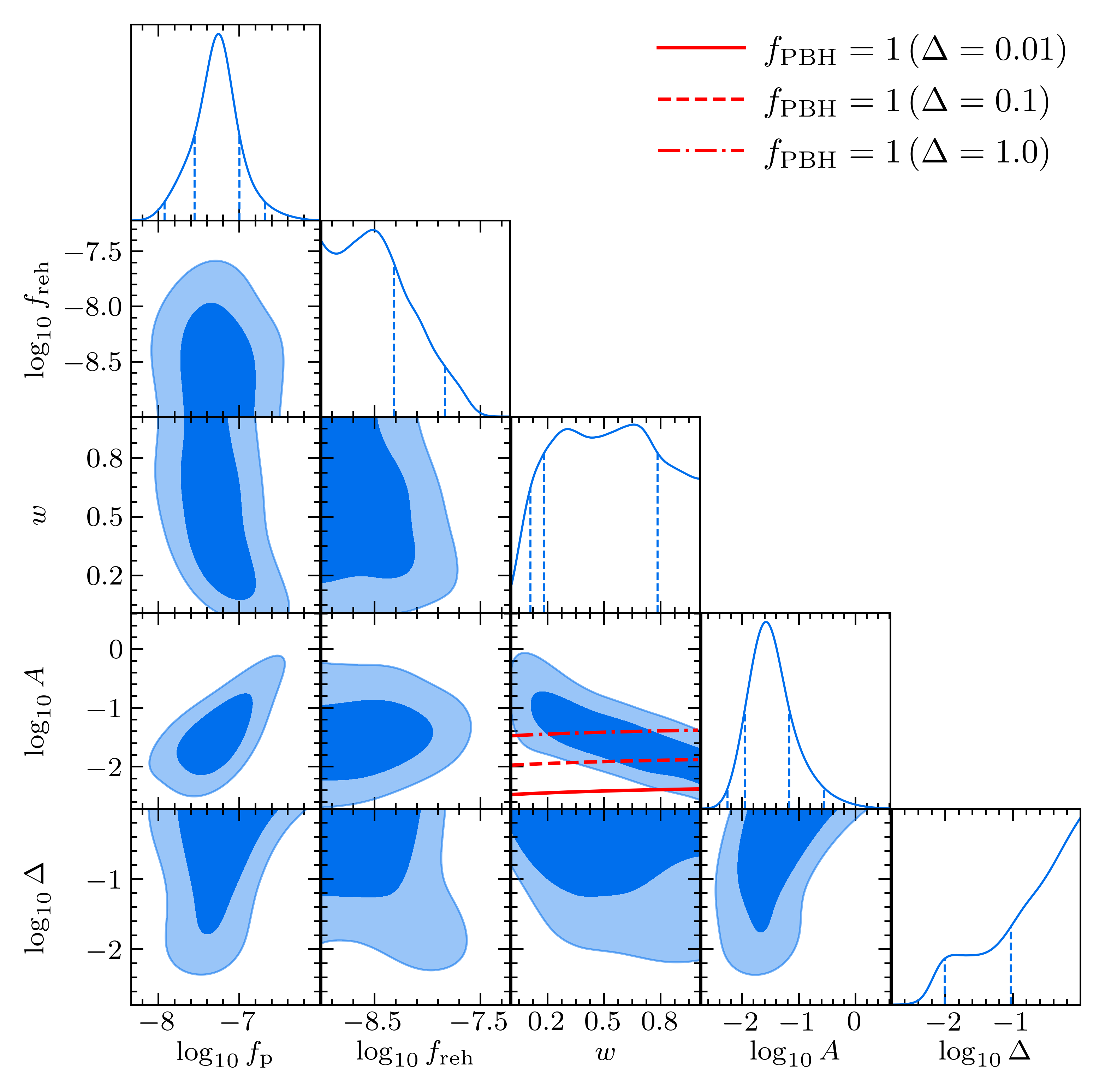} 
\caption{Same as Fig. \ref{fig:delta}, but with the width $\Delta$ as a new model parameter in the case of narrow log-normal spectrum. In the joint posterior distribution of $w$ and $A$, the solid, dashed, and dash-dotted red lines correspond to the PBH abundance $f_{\rm PBH} = 1$, with $\Delta = 0.01$, $0.1$, and $1$, respectively, and only the regions below these lines can avoid PBH overproduction. As $\Delta$ increases, the parameter space allowed for both the PTA and PBH constraints becomes larger accordingly.} \label{fig:narrow}
\end{center}
\end{figure*}

Now, we turn to the central part of our work: the case of a narrow log-normal spectrum. We mean ``narrow'' by setting the prior of the width $\Delta$ of primordial power spectrum as $\log_{10}\Delta\in [-3,0]$. The posterior distributions of all five model parameters (i.e., the same four as those in Sec. \ref{sec:delta} and $\Delta$) are shown in Fig. \ref{fig:narrow}. 

First, we observe that both $f_{\rm p}$ and $f_{\rm reh}$ are more tightly constrained than the case of the Dirac $\delta$ spectrum ($f_{\rm p}\sim 10^{-8}$--$10^{-7}$ Hz and $f_{\rm reh}$ is primarily localized in a low-frequency region around $10^{-9}$--$10^{-8}$ Hz). In addition to the necessary condition $f_{\rm reh}<f_{\rm p}$, there exhibits a correlation between them, as can be seen from their joint posterior distribution. Physically speaking, this indicates that the energy scale at the end of reheating cannot be too high. Otherwise, the spectral feature dominated by the reheating phase will deviate from the PTA band, rendering the characteristic cutoff unable to be fitted, so the cutoff feature provides a theoretical upper bound on the end time of cosmic reheating \cite{Domenech:2020kqm, Domenech:2024rks}. 

\begin{figure}[htbp]
\centering\includegraphics[width=1\linewidth]{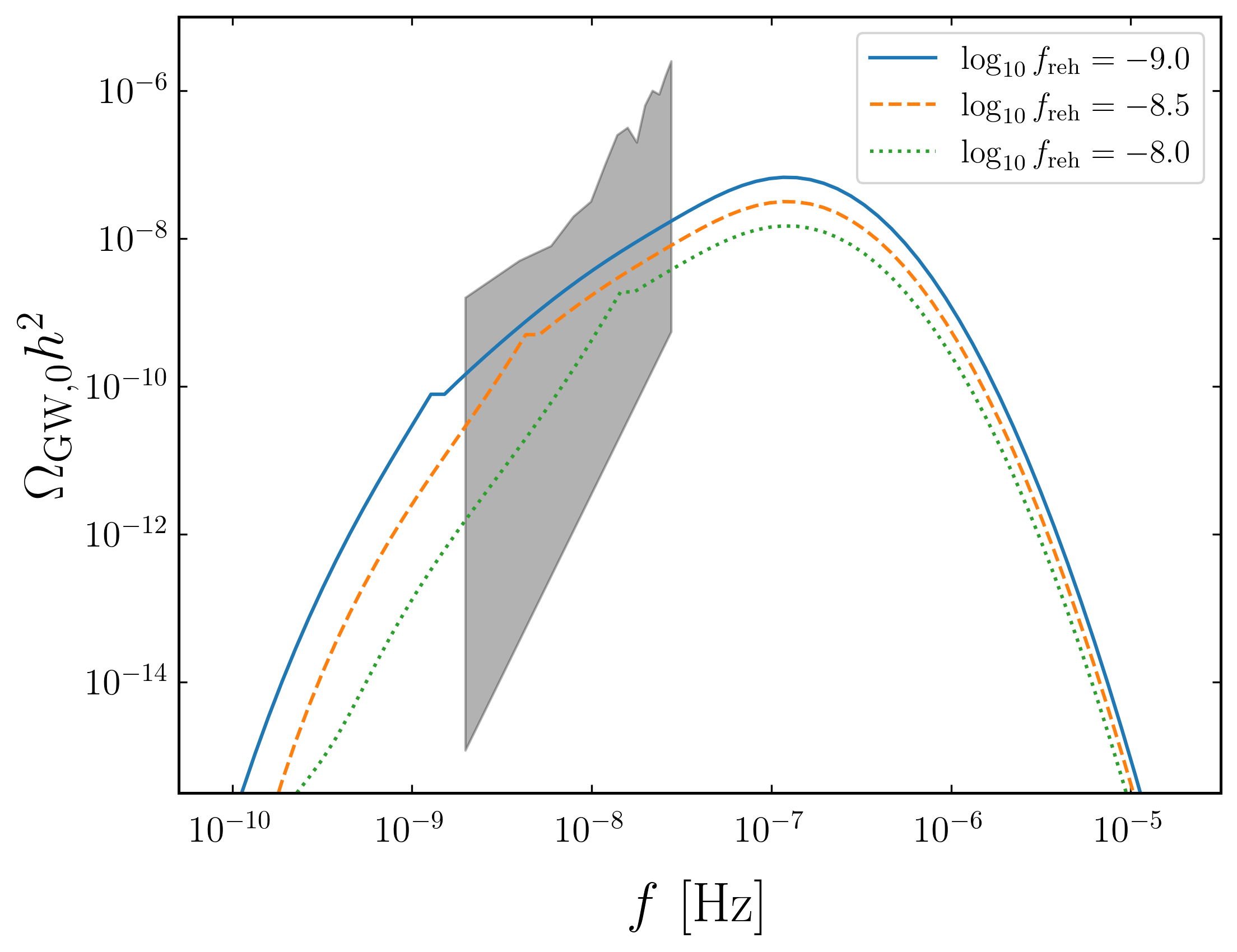}
\caption{The SIGW spectra for different reheating frequencies $\log_{10} f_{\rm reh}=-9$, $-8.5$, and $-8$, respectively, with other model parameters chosen as their maximum marginal posterior values ($\log_{10} f_{\rm p}=-7.25$, $w=0.66$, $\log_{10} A=-1.57$, and $\log_{10} \Delta=0$). The gray shaded region is favored by the NANOGrav 15-yr dataset. It can be clearly seen that, with the increase of $f_{\rm reh}$, the position of the characteristic frequency knee in the infrared tail in the SIGW spectrum shifts to higher frequencies.} \label{fig:freh_effect}
\end{figure}

\begin{figure}[htbp]
\centering\includegraphics[width=1\linewidth]{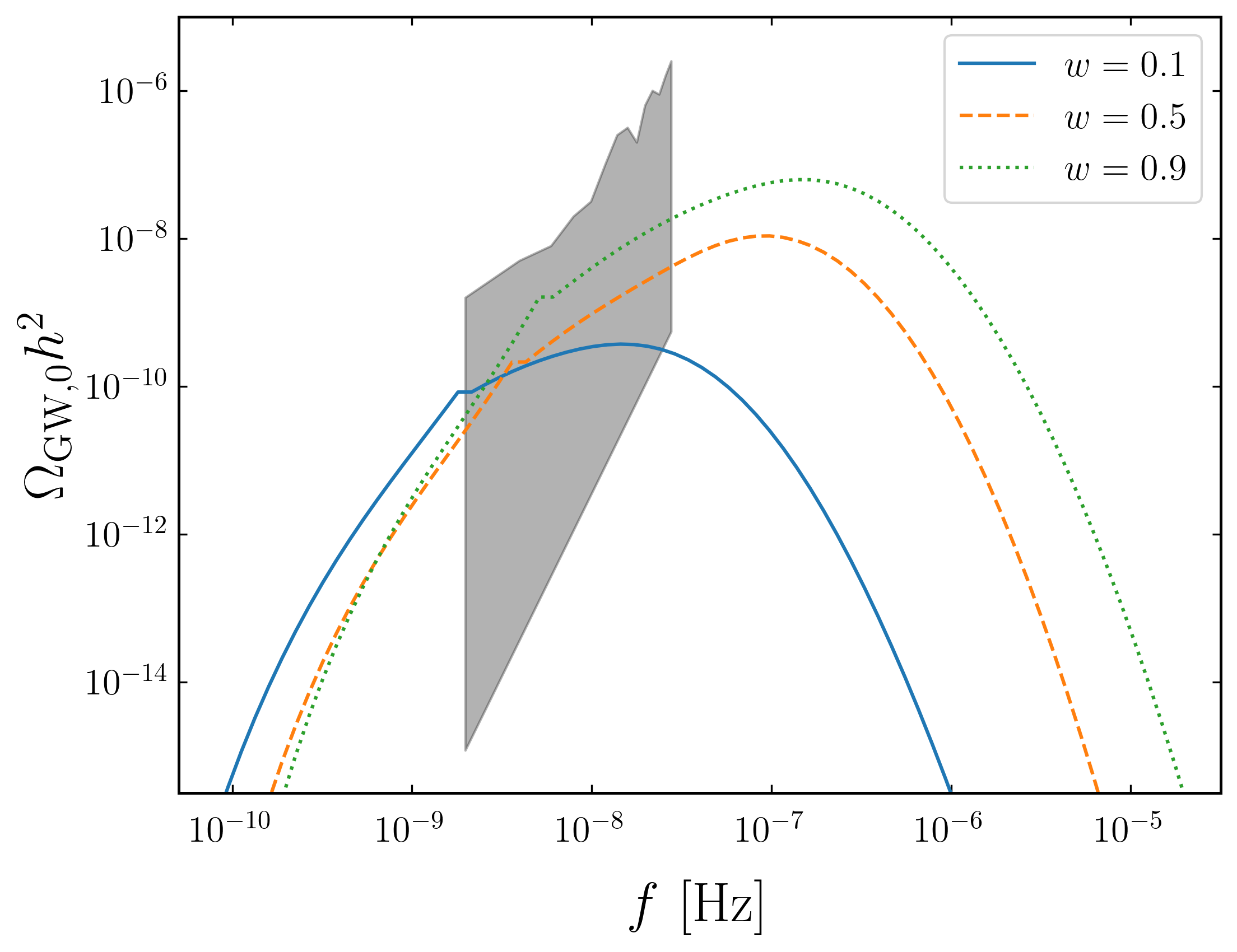}
\caption{Same as Fig. \ref{fig:freh_effect}, but with different EOS parameters $w=0.1$, $0.5$, and $0.9$, respectively ($\log_{10} f_{\rm p}=-7.25$, $\log_{10} f_{\rm reh}=-8.51$, $\log_{10} A=-1.57$, and $\log_{10} \Delta=0$). It can be found that a stiffer cosmic background with larger values of $w$ results in a steeper intermediate slope between the characteristic frequency knee and the peak in the SIGW spectrum, and also significantly enhances the peak amplitude.} \label{fig:w_effect}
\end{figure}

Second, at the $68\%$ CL, the constraint on the EOS parameter $w$ reads $w = 0.53^{+0.26}_{-0.34}$, consistent with joint-array analyses. For example, a recent study combining the NANOGrav, EPTA, and PPTA data reported a median value of $w \approx 0.60$ under a strict narrow-peak limit ($\Delta \le 0.1$) \cite{Liu:2023pau}. However, we should emphasize that, compared to previous works, in which either $f_{\rm reh}$ was fixed \textit{a priori} \cite{Domenech:2024rks} or the thermal scale was implicitly restricted by fixing the spectral shapes \cite{Liu:2023pau}, our current analysis treats $f_{\rm reh}$ as a completely free parameter in the full parameter space. This reveals a profound physical degeneracy: albeit $f_{\rm p}$ is constrained in a relatively narrow region, the low-frequency asymptotic slope of the SIGW spectrum within the observational band is determined by $f_{\rm reh}$ and $w$ together (to be seen in Figs. \ref{fig:freh_effect} and \ref{fig:w_effect}). 
This prevents an independent and high-precision determination of $w$ merely from PTA data, also consistent with the relation between $\Omega_{\rm GW,0}$ and $(k/k_{\mathrm{reh}})^{-2\beta}$, as explained in the Dirac $\delta$ spectrum case.

Same as Sec. \ref{sec:delta}, we again discuss the impact of PBH overproduction, but this time we focus on the influence from the spectral width $\Delta$. In the joint posterior distribution of $w$ and $A$ in Fig. \ref{fig:narrow}, three lines illustrate the critical bound with $f_{\rm PBH} = 1$, with different values of $\Delta=0.01$, $0.1$, and $1$, respectively, and only the regions below these lines are physically allowed. When $\Delta$ is rather small, the red line is also located at the lower right corner, and almost all parameter space is disfavored, exactly resembling the case of the Dirac $\delta$ spectrum as expected (see Fig. \ref{fig:delta}). However, as $\Delta$ increases, the constraints from $f_{\rm PBH}$ are greatly relieved. For instance, when $\Delta=1$, half of the parameter space is allowed (at both the $68\%$ and $95\%$ CLs). Consequently, adding a new model parameter (i.e., a relatively broader width $\Delta$, but still in the narrow log-normal spectrum model) naturally alleviates the tension between fitting PTA data and avoiding PBH overproduction.

Furthermore, to intuitively comprehend the influences from the cosmic reheating history, we illustrate the SIGW spectrum $\Omega_{\rm GW, 0}h^2$ in Figs. \ref{fig:freh_effect} and \ref{fig:w_effect}, with different values of $f_{\rm reh}$ and $w$. (1) The end of the reheating phase imprints a characteristic frequency knee in the infrared tail in the SIGW spectrum. As $f_{\rm reh}$ is determined by the energy scale of reheating, a higher $f_{\rm reh}$ corresponds to an earlier end of reheating, thereby shifting the spectral knee toward higher frequencies, as shown in Fig. \ref{fig:freh_effect}. (2) The value of $w$ governs the background expansion rate, which significantly modifies the intermediate spectral shape and peak amplitude of the SIGW spectrum. From Fig. \ref{fig:w_effect}, a stiffer cosmic background ($w > 1/3$) yields a steeper slope between the frequency knee and the peak, while explicitly enhancing the overall SIGW energy density. Conversely, a softer EOS ($w < 1/3$) leads to a smaller slope and a relative suppression of the peak amplitude. 

\begin{figure}[htbp]
\begin{center}
\includegraphics[width=1\linewidth]{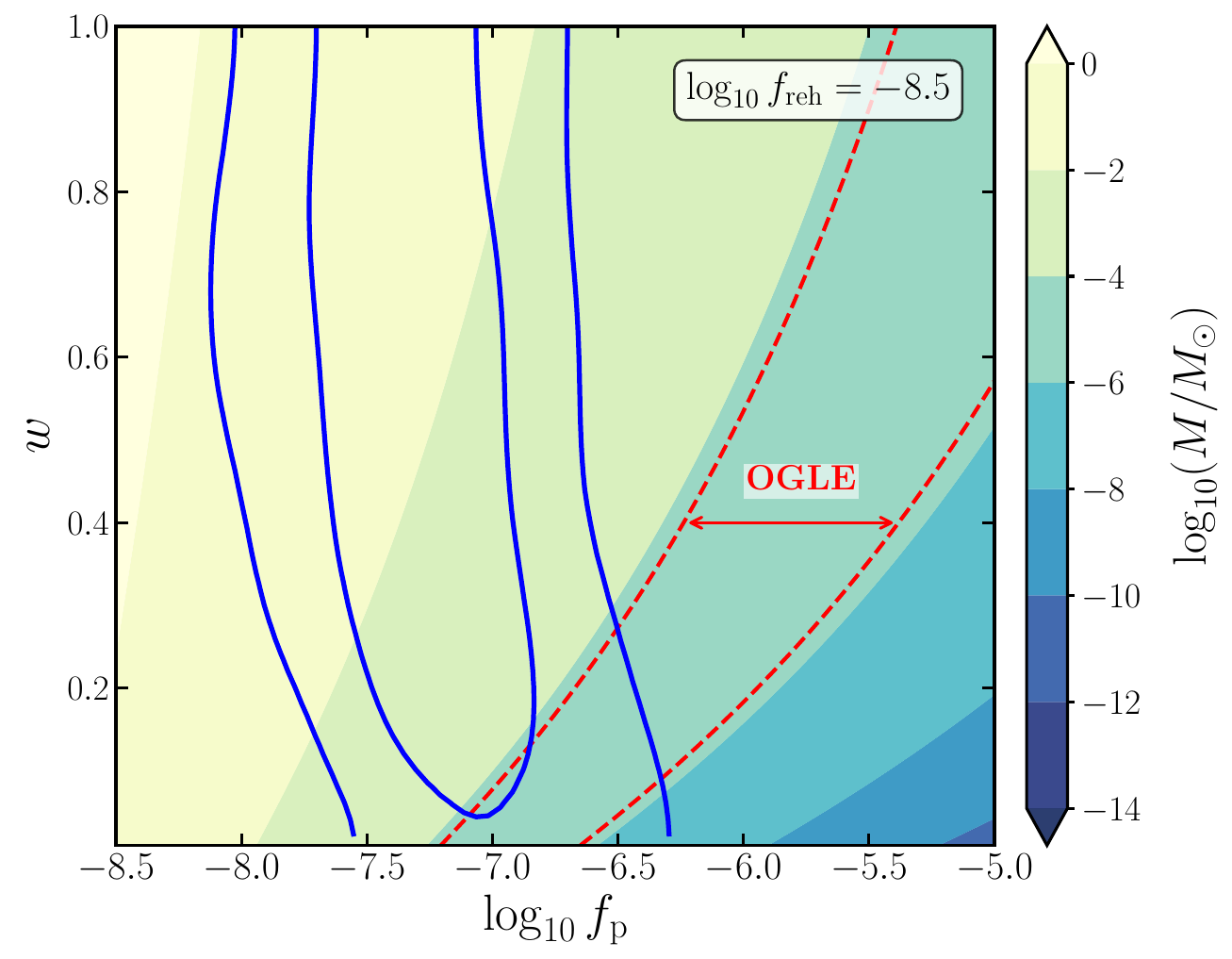} 
\caption{Same as Fig. \ref{fig:delta_pbh}, but for the case of the narrow log-normal spectrum, with $\log_{10}f_{\mathrm{reh}}=-8.5$. The considerably smaller overlap between the blue contours and the red region indicates that a finite spectral width significantly decreases the allowed parameter window, but it still leaves a specific narrow mass window for the PBHs to simultaneously satisfy the SIGW and microlensing observations. (We note that the theoretical PBH mass mapping for the finite-width spectra here is evaluated at a reference value of $\Delta \approx 0.5$.)} \label{fig:narrow_pbh}
\end{center}
\end{figure}

Last, we again explore the parameter degeneracy between $w$ and $f_{\rm p}$ in Fig. \ref{fig:narrow_pbh}. Similar to Fig. \ref{fig:delta_pbh}, the parameter space favored by PTA data is indicated by the blue lines. However, compared to the case of the Dirac $\delta$ spectrum, the overlap with the PBH mass range constrained by the OGLE events becomes much smaller. This considerably reduced overlap implies that introducing a finite spectral width places more stringent constraints on the model. Albeit the allowed parameter window severely shrinks, a consistent physical framework can still be established within this narrow window to simultaneously satisfy the PTA observations and the OGLE constraints.

\begin{figure*}[htbp]
\begin{center}
\includegraphics[width=0.85\linewidth]{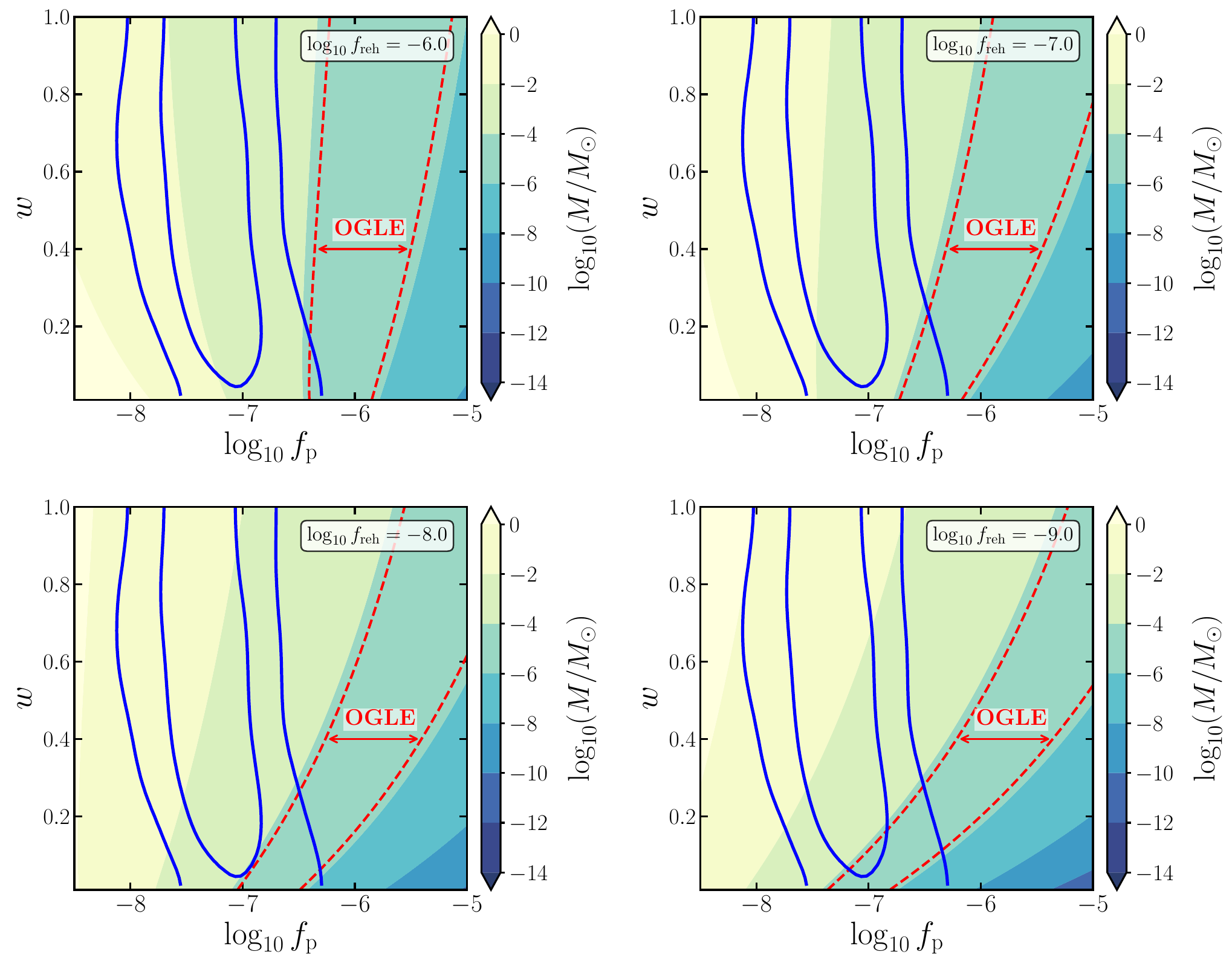}
\caption{The PBH mass mappings with varying reheating frequencies $f_{\rm reh}=10^{-9}$, $10^{-8}$, $10^{-7}$, and $10^{-6}$ Hz, respectively. Decreasing $f_{\rm reh}$ drives the PTA-favored parameter spaces toward larger mass scales, gradually overlapping the OGLE mass window. This observation reveals the parameter degeneracy: without independent constraint on the cosmic thermal history, the predicted PBH mass range can be adjusted to align with the OGLE events by tuning $f_{\rm reh}$.} \label{fig:compare}
\end{center}
\end{figure*}

Furthermore, Fig. \ref{fig:compare} illustrates the impacts from varying the reheating frequency $f_{\rm reh}$ on the PBH mass mappings. For fixed $f_{\rm p}$ and $w$, lowering $f_{\rm reh}$ (i.e., delaying the end of reheating) systematically shifts the PBH masses toward larger values, confirming $f_{\rm reh}$ as a dominant tuning parameter. Its variation changes the relevant mass range, directly telling whether the model predicts the $10^{-6}$--$10^{-4} M_\odot$ OGLE microlensing window. Thus, determining the reheating energy scale is a prerequisite for reliably linking the early-Universe GWs to macroscopic DM observations.

\subsection{Broad log-normal spectrum} \label{sec:broad} 

\begin{figure*}[htbp]
\begin{center}
\includegraphics[width=0.65\linewidth]{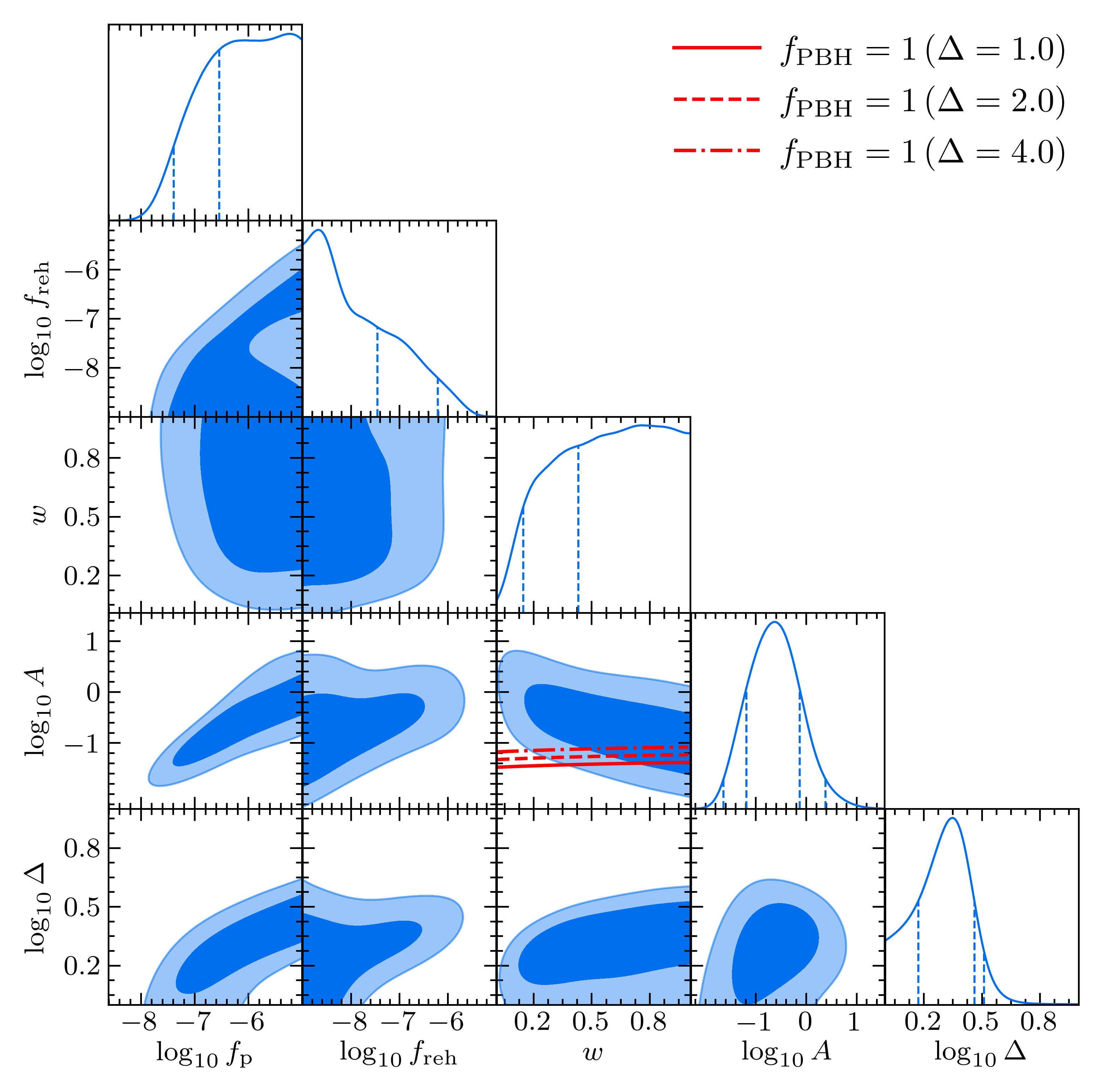}
\caption{Same as Fig. \ref{fig:narrow}, but in the case of the broad log-normal power spectrum, with $\Delta=1$, 2, and 4, respectively. The posterior distributions of $f_{\rm p}$, $f_{\rm reh}$, and $w$ are more diffusive compared to those in Fig. \ref{fig:narrow}.} \label{fig:wide}
\end{center}
\end{figure*}

Finally, we complete our work by investigating the case of a broad log-normal spectrum, with the width $\Delta>1$. The posterior distributions of the five model parameters are shown in Fig. \ref{fig:wide}. The relevant discussions are parallel to those in Sec. \ref{sec:narrow}, so we only need to focus on their discrepancies. 

\begin{figure}[htbp]
\centering\includegraphics[width=1\linewidth]{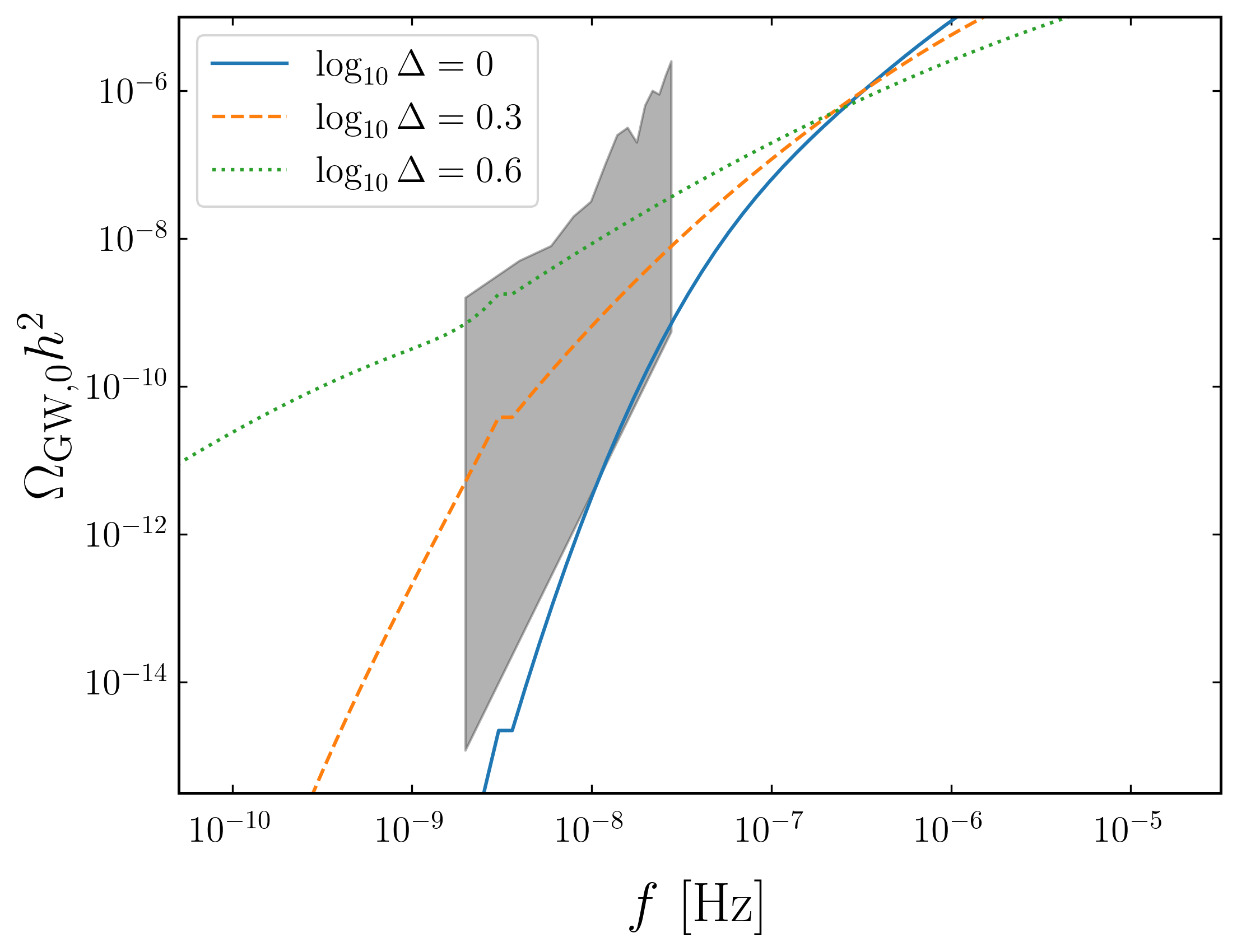}
\caption{Same as Fig. \ref{fig:freh_effect}, but with different spectral widths $\log_{10} \Delta=0$, $0.3$, and $0.6$, respectively ($\log_{10} f_{\rm p}=-5.23$, $\log_{10} f_{\rm reh}=-8.67$, $w=0.76$, and $\log_{10} A=-0.63$). As $\Delta$ increases, the SIGW spectra become significantly wider and flatter, smearing out the subtle features associated with the reheating history.} \label{fig:delta_effect}
\end{figure}

First, generally speaking, in contrast to the narrow spectrum model, with the spectral width $\Delta$ increasing, the constraining power of the broad spectrum model on the physical parameters is substantially weakened. The superposition of broad momentum modes will induce a ``smoothing effect'' on the SIGW spectra, making them wider and flatter, as shown in Fig. \ref{fig:delta_effect}. Crucially, this smoothing effect smears out the characteristic frequency knee and the subtle infrared slope variations induced by the reheating history (compared with Figs. \ref{fig:freh_effect} and \ref{fig:w_effect}). As a result, the reheating frequency $f_{\rm reh}$ can no longer be effectively constrained, and its posterior distribution exhibits a diffusive behavior in Fig. \ref{fig:wide}. This implies that, as long as the primordial power spectrum is broad enough, the reheating processes ending at different energy scales can all fit the existing PTA data fairly well.

Second, Bayesian inference reveals that the broad spectrum model has a relatively weak inference power for the EOS parameter $w$. The posterior distribution of $w$ in Fig. \ref{fig:wide} effectively rules out the matter-dominated Universe ($w \to 0$), but it flattens as a plateau when $w \gtrsim 0.3$. This observation indicates that, although the broad spectrum model excludes a soft EOS, it still lacks the resolving power to distinguish various stiffer phases, such as the RD ($w=1/3$) and kinetic-energy-dominated ($w=1$) cases. 
Our result confirms the analysis in Ref. \cite{Domenech:2024rks} (with a fixed $f_{\rm reh}$), which claimed that a broad spectrum yields the SIGW spectra with the shapes insensitive to $w$. 

Consequently, the signatures in the SIGW spectrum required to break the parameter degeneracy are nearly erased, due to the broad spectral width $\Delta$. This explains why PTA data fail to provide robust bounds on $f_{\rm reh}$ and $w$,
and the increasing $\Delta$ acts as a filter that hides the thermal history of the early Universe.

Furthermore, regarding the PBH abundance constraints, as $\Delta$ increases, the allowed parameter space becomes larger in the joint posterior distribution of $w$ and $A$ in Fig. \ref{fig:wide}, similar to that in Fig. \ref{fig:narrow}. Therefore, if $\Delta$ is large enough, 
the issue of PBH overproduction becomes unimportant, and most model parameter combinations that can successfully explain the PTA signals naturally satisfy the PBH constraints. This lack of constraining power, combined with the vast parameter uncertainty, invalidates a quantitative evaluation of the PBH fraction as DM.

Last, as the broad spectrum model allows $\Delta$ to vary over a large range, it causes the actual PBH mass corresponding to the posterior parameters to deviate significantly from the background mapping. This indicates that, under broad spectrum conditions, the PBH mass predicted by the model is actually uncertain. Without definite prior information on the thermal history of reheating, the broad spectrum model cannot provide physically meaningful constraints on the PBH mass, so we will not show the similar plot like Figs. \ref{fig:delta_pbh} and \ref{fig:narrow_pbh} any more.

\section{Conclusion} \label{sec:con}

The cosmic reheating is an important epoch in the very early Universe. The relevant energy scale and the constituents of cosmic medium can be encoded in two parameters: the reheating frequency $f_{\rm reh}$ and the EOS parameter $w$. In this work, we wish to explore the impacts of the thermal history of cosmic reheating on the PBH and SIGW physics. First, the PBH mass, abundance, and the SIGW spectrum are reviewed with a general $w$. Then, we utilize the NANOGrav 15-yr dataset and apply the full parameter space Bayesian inference approach, so as to constrain the primordial power spectrum ${\cal P}_{\cal R}(k)$, with both $f_{\rm reh}$ and $w$ taken as free model parameters. Three different types of ${\cal P}_{\cal R}(k)$: the Dirac $\delta$, narrow and broad log-normal spectra, are systematically investigated, with the spectral width $\Delta$ increasing in order. The issue of PBH overproduction and the comparison with the OGLE microlensing events are also carefully considered, which further shrink the allowed parameter spaces. Our study not only validates the capacity of PTA data to probe the primordial cosmological fluctuations, but also helps unveil the intrinsic correlations between reheating history and GW and PBH physics. Our main conclusions are summarized as follows.

First, the form of the primordial power spectrum ${\cal P}_{\cal R}(k)$ (in particular, its spectral width $\Delta$) plays an important role in reconstructing the cosmic thermal history. Our analysis indicates that the NANOGrav 15-yr dataset can effectively constrain the peak frequency $f_{\rm p}$ of the SIGW spectrum, when ${\cal P}_{\cal R}(k)$ possesses a relatively narrow width $\Delta\lesssim 1 $. On the contrary, a broader ${\cal P}_{\cal R}(k)$ usually induces a smoothing effect that obscures the characteristic frequency knee in the SIGW spectrum, thus losing the information of cosmic reheating. 

Second, there exists a parameter degeneracy between the reheating frequency $f_{\rm reh}$ and the EOS parameter $w$. Once $f_{\rm reh}$ is treated as a free model parameter, the shift of the spectral knee location strongly couples to the variation of the spectral slope driven by $w$, significantly decreasing the precision of constraints. This coupling causes the current PTA-based stochastic GW background data to fall short of independent ability to distinguish whether the Universe underwent a matter-dominated or kinetic-energy-dominated phase during reheating. 

Third, the specific spectral width assumptions may establish a bridge between PTA signals and microlensing events. In the Dirac $\delta$ and narrow log-normal spectrum models, the PBH masses associated with the SIGWs are uniquely anchored at the planetary mass scale (around $10^{-5} M_\odot$). This mass window is highly consistent with the DM candidates implied by the recent OGLE observations. This implies that, if the nanohertz stochastic GW background is truly due to the SIGW from the primordial power spectrum with a finite width, the Universe must host a vast population of PBHs with planetary masses. Such PBHs not only naturally explain the OGLE results, but may even constitute a significant proportion of DM. Fully verifying this comprehensive framework necessitates future research, combined with precise statistical distributions of microlensing events, to constrain the PBH abundance more strictly.

Last, we should again emphasize that the above results are obtained under the instantaneous transition approximation. 
The finite-duration transition would mainly smooth the spectral knee and thus introduce systematic uncertainty into the fitted effective reheating scale \cite{Inomata:2019zqy, Inomata:2019ivs},
while the limiting behaviors of the modes reentering well before or after reheating are expected to remain qualitatively unchanged. A complete treatment with a continuous EOS parameter $w(a)$ and a model-dependent transfer function requires specifying a concrete reheating scenario and is left for our future work. 

In conclusion, PBHs and SIGWs are becoming highly potential cosmological probes. Albeit the current PTA data, limited by frequency bandwidth and observation precision, struggle to completely break the degeneracies between the thermal history parameters without imposed priors, this points the way forward for future multiband GW astronomy. With the gradual release of next-generation space-based GW detectors (e.g., LISA \cite{Barke:2014lsa}, Taiji \cite{Ruan:2018tsw}, and TianQin \cite{Gong:2021gvw}, as well as the ground-based detectors \cite{Evans:2021gyd}) and higher-precision PTA data, capturing the broken power-law features in the SIGW spectra across much broader frequency bands will be the ultimate path to decoding the thermal history in the very early Universe. Breaking these degeneracies permanently not only requires the cosmic microwave background observations to constrain the initial reheating conditions, but also signifies the dawn of a new era of multimessenger DM astronomy, integrating GW detection with electromagnetic observations like microlensing \cite{Mingarelli:2026pur, Liu:2026tik, Li:2026pjy, Sardana:2026jvr, Ghaleb:2025xqn, Allen:2025waa, KAGRA:2022osp}.

\begin{acknowledgments}
We are very grateful to Xun Chen, Guillem Dom\`{e}nech, Shi Pi, Dominik J. Schwarz, Bing-Yu Su, and Yu-Jia Zhao for fruitful discussions. This work is supported by the National Natural Science Foundation of Liaoning Province, China (No. 2025-MS-034).
\end{acknowledgments}

\section*{DATA AVAILABILITY}
The data that support the findings of this article are openly available.


\bibliography{apssamp}
\end{document}